\documentclass[aps,article,preprint,superscriptaddress,showpacs,amsmath,amssymb,
amsfonts,balancelastpage]{revtex4-1}
\usepackage{amsmath}
\usepackage{amssymb}
\usepackage{amsfonts}
\usepackage[justification=raggedright]{caption}
\usepackage{subcaption}
\usepackage{graphicx}
\usepackage{epsfig}
\usepackage{hyperref}
\usepackage{color}
\usepackage{color, soul}
\pdfoutput=1
\usepackage{float}
\usepackage{xcolor}
\hypersetup{
colorlinks=true,
urlcolor=blue,
citecolor=blue,
linkcolor=blue,}

\begin{document}
\title{Evidence of a pseudogap driven by competing orders of multi-band origin in the ferromagnetic superconductor Sr$_{0.5}$Ce$_{0.5}$FBiS$_2$  }

\author{Mohammad Aslam}
\affiliation{Department of Physical Sciences,
Indian Institute of Science Education and Research Mohali,
Sector 81, S. A. S. Nagar, Manauli, PO: 140306, India}

\author{Arpita Paul }
\affiliation{Theoretical Sciences Unit, Jawaharlal Nehru Centre for Advanced Scientific Research, Jakkur, Bangalore 560 064, India}

\author{Gohil S. Thakur}
\affiliation{Department of Chemistry, Indian Institute of Technology, New Delhi 110016, India}
\author{Sirshendu Gayen}
\affiliation{Department of Physical Sciences,
Indian Institute of Science Education and Research Mohali,
Sector 81, S. A. S. Nagar, Manauli, PO: 140306, India}

\author{Ritesh Kumar}
\affiliation{Department of Physical Sciences,
Indian Institute of Science Education and Research Mohali,
Sector 81, S. A. S. Nagar, Manauli, PO: 140306, India}

\author{Avtar Singh}
\affiliation{Department of Physical Sciences,
Indian Institute of Science Education and Research Mohali,
Sector 81, S. A. S. Nagar, Manauli, PO: 140306, India}

\author{Shekhar Das}
\affiliation{Department of Physical Sciences,
Indian Institute of Science Education and Research Mohali,
Sector 81, S. A. S. Nagar, Manauli, PO: 140306, India}

\author{Ashok K. Ganguli}
\affiliation{Department of Chemistry, Indian Institute of Technology, New Delhi 110016, India}
\affiliation{Institute of Nano Science $\&$ Technology, Mohali 160064, India.}

\author{Umesh V. Waghmare}
\affiliation{Theoretical Sciences Unit, Jawaharlal Nehru Centre for Advanced Scientific Research, Jakkur, Bangalore 560 064, India}
\author{Goutam Sheet}
\affiliation{Department of Physical Sciences,
Indian Institute of Science Education and Research Mohali,
Sector 81, S. A. S. Nagar, Manauli, PO: 140306, India}
\date{\today}

\begin{abstract}

From temperature and magnetic field dependent point-contact spectroscopy on the ferromagnetic superconductor Sr$_{0.5}$Ce$_{0.5}$FBiS$_2$ (bulk superconducting $T_c$ = 2.5 K) we observe  (a) a pseudogap in the normal state that sustains to a remarkably high temperature of 40 K and (b) two-fold enhancement of $T_c$ upto 5 K in  the point-contact geometry. In addition, Andreev reflection spectroscopy reveals a superconducting gap of  6 meV for certain point-contacts suggesting that the mean field $T_c$ of this system could be approximately 40 K, the onset temperature of pseudo-gap. Our results suggest that quantum fluctuations originating from other competing orders in Sr$_{0.5}$Ce$_{0.5}$FBiS$_2$ forbid a global phase coherence at high temperatures thereby suppressing $T_c$. Apart from the known ordering to a ferromagnetic state, our first-principles calculations reveal nesting of a multi-band Fermi surface and a significant electron-phonon coupling that could result in charge density wave-like instabilities.
\end{abstract}
\maketitle
\textbf{1. Introduction:}\\
Within the ambit of the microscopic theory of superconductivity developed by Bardeen, Cooper and Schriefer (BCS) \cite{Bardeen, Berk}, superconductivity and ferromagnetism are two antagonistic phenomena \cite{Al, Ber}. While in the high temperature superconductors it is believed that the superconducting pairing is mediated by magnetic interactions \cite{n.d}, it is also known that superconductivity competes with magnetism. For example, in certain cuprate superconductors the critical temperature is enhanced when the antiferromagnetic ordering is suppressed \cite{r.e, m.s, yong}. On the other hand there are few systems where superconductivity is seen to co-exist with ferromagnetism \cite{Li, Ren, J.Lee, s.s, d}. Perhaps the most recently discovered member of the ferromagnetic superconductor family is Sr$_{0.5}$Ce$_{0.5}$FBiS$_2$ where the Ce ions are known to order ferromagnetically at 7.5 K and the system undergoes a superconducting transition below 3 K.

Sr$_{0.5}$Ce$_{0.5}$FBiS$_2$ is derived by doping Ce in the Sr sites of the parent compound SrFBiS$_2$, which is a semiconductor. BiS$_2$ based superconductors in general have attracted tremendous interest in recent times due to their remarkable structural similarities with the high T$_c$ cuprate and the ferropnictide supercondcutors \cite{Jiangping, Xing, Usui, Yild}. Like in the cuprates and the ferropnictides, the BiS$_2$ based superconductors posses layered crystal structure where the BiS$_2$ layers superconduct, and variety of superconductors are obtained by changing the intercalating block layers \cite{Yoshkazu, Xi, Sugi, Zeng, H, Y, S}. Primarily because of this similarity, significant interest has developed in studying the electronic structure, the pairing interactions and the magnetic interactions in the BiS$_2$ based superconductors.

In this Letter, from detailed temperature and magnetic field dependent point-contact spectroscopy measurements, we show that superconductivity in the ferromagnetic superconductor Sr$_{0.5}$Ce$_{0.5}$FBiS$_2$ competes with other existing orders thereby leading to a normal state pseudogap. Using first principles calculations, we show the evidence of nesting in the multi-band fermi surface of this material \cite{M.d, wan, Lamura, Ye, Baumb, G.B, Lee}. From the detailed analysis of the experimental and the theoretical results we conclude that the pseudogap in Sr$_{0.5}$Ce$_{0.5}$FBiS$_2$ originates from fluctuations of the phase of the complex superconducting order parameter. These fluctuations probably result from the quantum phase transitions to other competing orders supported by the multi-band Fermi surface of Sr$_{0.5}$Ce$_{0.5}$FBiS$_2$. This idea is further supported by the observation of a two-fold enhancement of the local $T_c$ when the nesting becomes weaker at metallic point-contacts formed on Sr$_{0.5}$Ce$_{0.5}$FBiS$_2$ .

\textbf{2. Experimental results and  discussion:}\\
The point-contact spectroscopy experiments were performed at low temperatures in a liquid helium based cryostat. All the measurements presented in this Letter were performed on a polycrystalline sample of Sr$_{0.5}$Ce$_{0.5}$FBiS$_2$  and the metallic point-contacts on the sample were fabricated by sharp tips of pure palladium (Pd). Statistically we found two types of $dI/dV$ spectra obtained at different points on polycrystalline Sr$_{0.5}$Ce$_{0.5}$FBiS$_2$. Two such representative spectra are shown in Figure 1(a) (spectrum \textbf{type A}) and Figure 1(b) (spectrum \textbf{type B}) respectively. It should be noted that the observation of two types of spectra might also be possible due to the existence of phase inhomogeneity in the sample. Within the standard characterization tools (including X-ray diffraction analysis) that we have used to determine the phase purity of the samples, all the samples formed in single phase. On the other hand, on an average, since the two types of spectra appeared with equal number of times, an inhomogeneous phase would consist of two phases distributed over the surface in almost equal proportion. However, in that case the two phases would be distinctly identified in XRD analysis. Therefore, the emergence of two types of spectra due to phase inhomogeneity can be ruled out. 
\begin{figure}[h!]
	\centering
		\includegraphics[width=\textwidth]{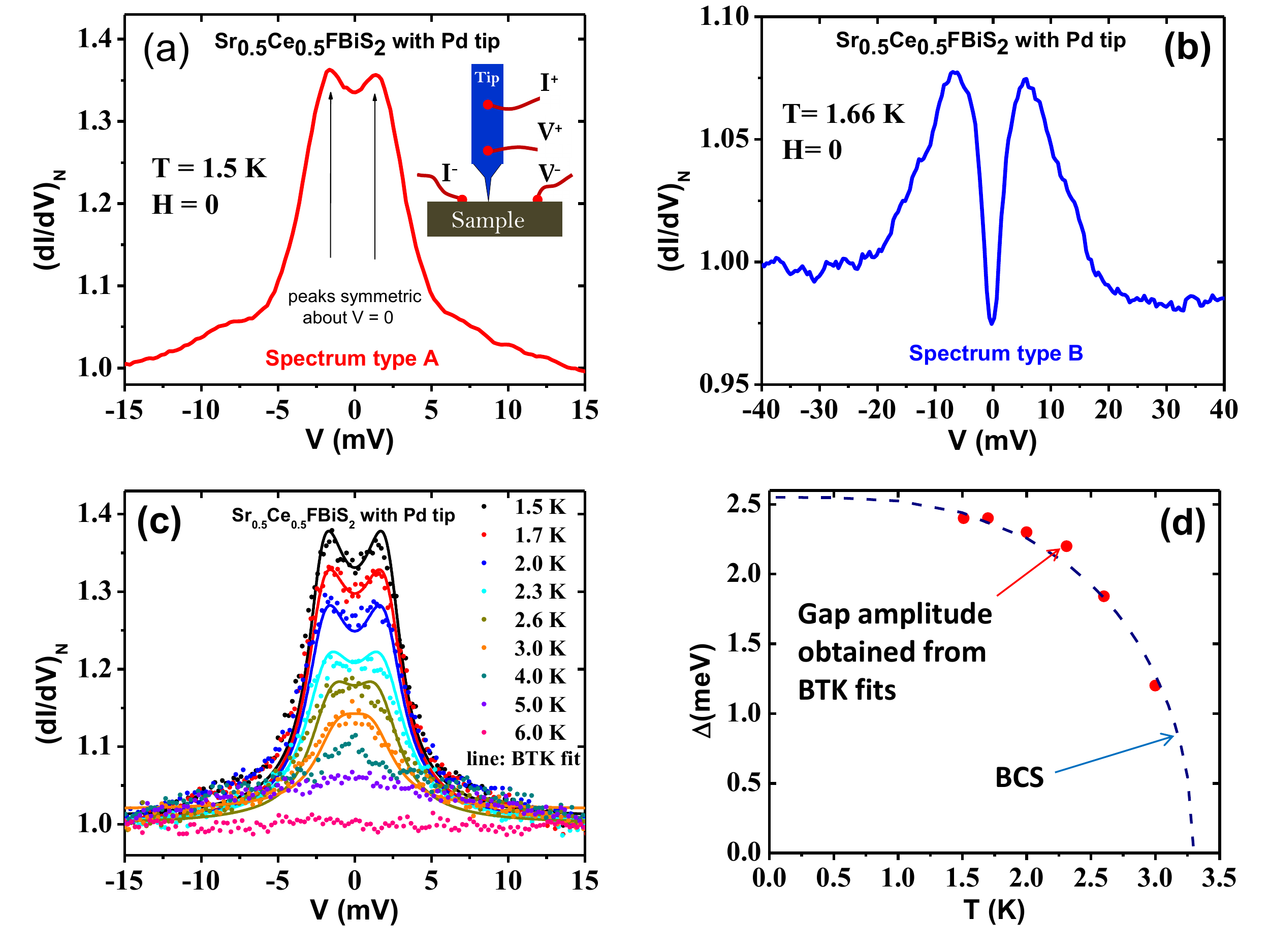}
	\caption{ \textbf{Different types of point-contact spectra obtained on Sr$_{0.5}$Ce$_{0.5}$FBiS$_2$} (a) A representative spectrum of \textbf{type A} ($\Delta$ $\simeq$ $2.4$ meV). The double peaks structure symmetric about $V = 0$ indicated by arrows is a hallmark of Andreev reflection in NS point-contacts. (b) A representative spectrum of \textbf{type B} ($\Delta$ $\simeq$ $6$ meV). (c) Temperature dependence of a spectrum of \textbf{type A} with BTK fits (indicated by solid lines) of each spectra. (d) Temperature dependence of $\Delta$ extracted from BTK analysis of the temperature dependent spectra. The dotted line shows the BCS prediction.}
	\label{Figure 1}
\end{figure}
 The spectrum shown in Figure 1(a) has a double peak structure symmetric about $V = 0$. Such a double peak structure is a hallmark of Andreev reflection in ballistic normal metal-superconductor (NS) point-contacts. The position of the  peaks in such spectra gives an estimate of the amplitude of the superconducting energy gap. Fitting this spectrum with the theoretical model of Blonder, Tinkham and Klapwijk (BTK), our estimate of the superconducting energy gap is $\Delta$ $\simeq$ 2.4 meV.  In the later part of this Letter, we will refer to this type of spectrum as \textbf{type A}. The spectrum shown in Figure 1(b) also shows the double peak structure symmetric about $V=0$. However, the spectrum is significantly broadened and consequently the spectrum deviates from BTK-like behavior. If the position of the peaks is taken as a rough estimate of the superconducting energy gap for such a spectrum, the gap amplitude turns out to be  $\Delta$ $\simeq$ 6 meV. In the later part of this Letter we will refer to this type of spectrum as \textbf{type B}. We surmise that the observation of two types of spectra with prominent anisotropic features originate from the anisotropic band structure of Sr$_{0.5}$Ce$_{0.5}$FBiS$_2$ (confirmed by calculated band structure presented later).

We have investigated the temperature dependent behavior of the superconducting energy gap as extracted from the \textbf{type A} spectrum. For this, we recorded the spectra at different temperatures and fitted the spectra using BTK formalism. The temperature dependent spectrum as well as the BTK fitting are shown in Figure 1(c) as dotted and solid lines respectively. It is seen that the features associated with Andreev reflection vary smoothly with temperature and the double peak structure disappear around 2.8 K. Beyond this temperature the low-bias conductance enhancement is present which decays smoothly with further increasing temperature and eventually all prominent spectral features associated with superconductivity disappear around 5 K. This temperature is approximately 2 times higher than the bulk $T_c$. It is important to note that the spectra could be fitted nicely using the BTK model which was developed for conventional BCS superconductors. Furthermore, the extracted temperature dependence of the superconducting energy gap shown in Figure 1(d) follows the trend as expected in BCS theory provided the low-temperature gap is assumed to be 2.4 meV with a $T_c$ of 3.3 K. A gap amplitude of 2.4 meV indicates high value of $\Delta/{k_BT_c} \sim 12$, which is large even in comparison with the known strong coupling superconductors. In the other hand the energy gap ($\Delta$) extracted from \textbf{type B} spectra  is 6 meV. In fact, such a high value of the energy gap suggests a mean field critical temperature $T_c$ of around 40 K (2$\Delta$ $\simeq$ 3.5 K$_B$T$_C$) within the formalism of weak-coupling conventional BCS theory. In the past it was shown for certain unconventional \cite{Goutam, Davis} and disordered conventional superconductors \cite{Pratap} that superconducting correlations might emerge at a higher temperature while the systems did not superconduct down to a much lower temperature due to fluctuation of the phase of the complex superconducting order parameter \cite{Emery}. Following the above discussion, it is rational to surmise that in Sr$_{0.5}$Ce$_{0.5}$FBiS$_2$ local superconducting correlations emerge at temperatures as high as 40 K but there are other physical processes in Sr$_{0.5}$Ce$_{0.5}$FBiS$_2$ which make the phase of the order parameter fluctuate and does not allow a global phase-coherence until the system is cooled down to less than 3 K. Similarly, the gap corresponding to the spectra of \textbf{type A}  is around 2.4 meV for which the mean field $T_c$ should be around 16 K. This is also quite large considering the low $T_c$ of the superconductor. Since the system is close to a phase fluctuation regime, as evidenced by the features observed in the spectra type B, it is possible that the phase of the order parameter associated with the spectra of \textbf{type A} also fluctuates.

The magnetic field dependence of the spectrum of \textbf{type A} is shown in Figure 2(a) with respective BTK fits. Figure 2(b) shows a systematic evolution of the superconducting energy gap as a function of magnetic field. The upper critical field of the given point-contact is approximately 1.2 Tesla where all the spectral features disappear.
\begin{figure}[h!]
\centering
\includegraphics[width=\textwidth]{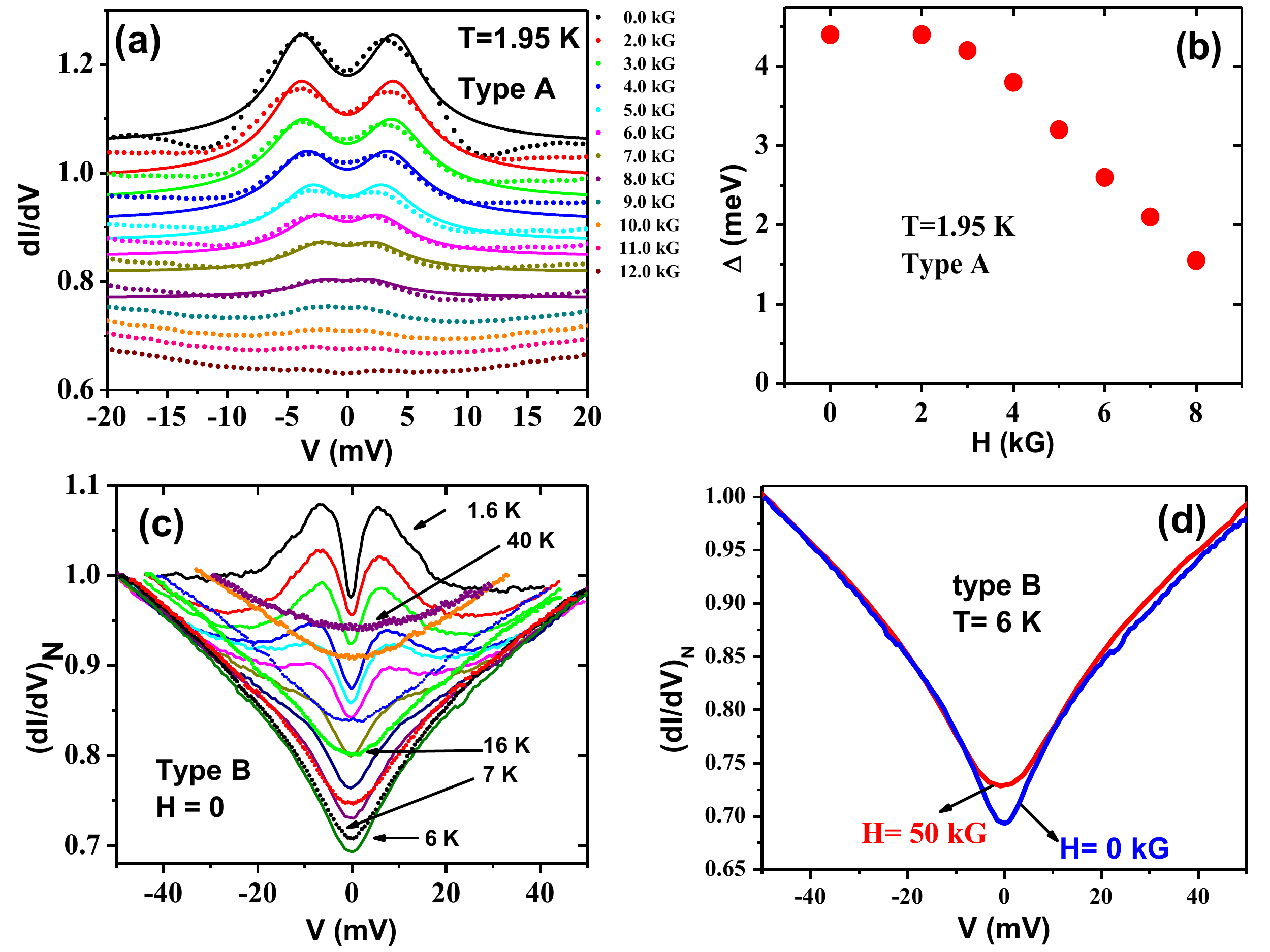}
\caption{(a) Magnetic field dependence of a representative spectra of \textbf{type A}. The solid lines show BTK fits. The spectra have been shifted vertically for visual clarity. (b) Magnetic field dependence of $\Delta$. (c) Temperature dependence of a representative spectrum of \textbf{type B}. (d) The spectrum of \textbf{type B} obtained at 6 K at 0 T and 5 T of magnetic field.}
\label{Figure 1}
\end{figure}

Now the natural question which must be addressed first is what physical mechanism might lead to the fluctuation of phase and suppression of superconductivity in Sr$_{0.5}$Ce$_{0.5}$FBiS$_{2}$ leading to the emergence of the pseudogap-like feature. To this end, we first consider the competing physical orders present in the system. From previous measurements it is known that in Sr$_{0.5}$Ce$_{0.5}$FBiS$_2$ a ferromagnetic order coexists with superconductivity. Proximity of ferromagnetic fluctuations may indeed suppress superconductivity leading to an effective $T_c$ much lower than the mean field $T_c$. However, the ferromagnetic transition of the system also happens at low temperature $\sim 7.5 K$. Therefore, it is important to identify other competing orders that might be responsible for dramatic reduction in $T_c$ from around 40 K to 3 K. In order to address this issue, we must now focus on the spectra of \textbf{type B} that is supposedly obtained when current is injected along a different momentum direction on the Fermi surface of polycrystalline Sr$_{0.5}$Ce$_{0.5}$FBiS$_2$ and the spectral features are predominantly governed by a different band crossing the Fermi surface.

The temperature evolution of the spectra of \textbf{type B} is shown in Figure 2(c). As mentioned earlier, these spectra indeed show the double peak structure as in the spectra of \textbf{type A}, but these spectra are significantly broadened and cannot be analyzed within the ambit of BTK theory. The temperature dependence of the overall spectral features reveals two associated energy scales. At lowest temperature the symmetric peaks in $dI/dV$ appear around  $\pm$ 6  meV respectively. This is one well defined energy scale in this case. The double dip structure is smoothly suppressed with increasing temperature which eventually becomes a strong dip at $V = 0$. This dip structure continues to exist above the global superconducting critical temperature of Sr$_{0.5}$Ce$_{0.5}$FBiS$_2$. This single dip structure becomes most intense at 6 K. Up to this temperature the spectra in Figure 2(c) are shown as solid lines. Beyond this temperature, the dip starts getting suppressed and such higher temperature spectra are represented by dotted lines for visual clarity. The suppression of the dip amplitude is rather slow and the dip eventually disappears in the background at a much higher temperature of 40 K. Beyond this temperature the spectrum remains temperature independent. This corresponds to the second energy scale in the system.

It should be noted that the dip structure above the superconducting $T_c$ changes only slightly with an applied magnetic field as high as 5 T. A sharp dip structure in differential conductance in the point-contact spectra between two non-superconducting metals (that do not show significant dependence on magnetic field but do show strong dependence with temperature) might also be related to a gap in the density of states associated with a charge density wave (CDW)-like phase originating due to the nesting of certain fermi surface pockets. Such a gap is seen for metallic point-contacts on NbSe$_3$, a standard example of a CDW-system \cite{sinchenko}. Therefore, the pseudogap structure observed here may be attributed to possible nesting in certain pockets of the Fermi surface of Sr$_{0.5}$Ce$_{0.5}$FBiS$_2$.

Having identified two distinct energy scales corresponding to two physical orders, it is now imperative to investigate the electronic structure of Sr$_{0.5}$Ce$_{0.5}$FBiS$_2$ and identify the bands relevant to various ordered phases and the coupling responsible for their stability. To address this issue, we have investigated the detailed electronic structure of bulk Sr$_{0.5}$Ce$_{0.5}$FBiS$_2$ and effects of its interaction with Pd tip using first-principles DFT-based calculations.

\begin{figure}[h!]
\centering
\includegraphics[width=\textwidth]{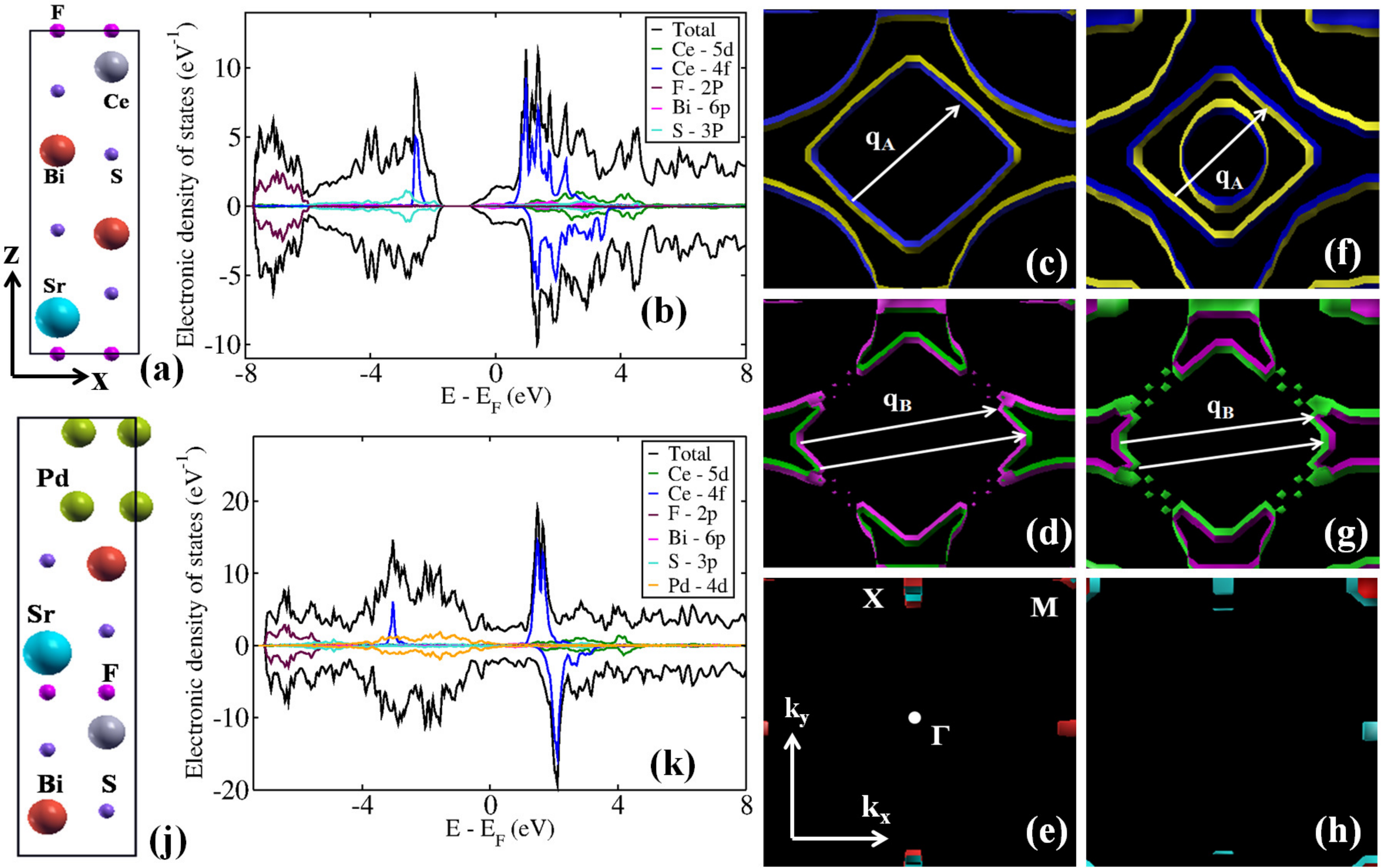}

\caption {(a) Crystal structure of bulk Sr$_{0.5}$Ce$_{0.5}$FBiS$_2$. (b) spin resolved electronic density of states of bulk Sr$_{0.5}$Ce$_{0.5}$FBiS$_2$. Spin resolved Fermi surface contributed by three bands crossing Fermi level: (c) band \textbf{A} for majority spin, (d) band \textbf{B} for majority spin, (e) band \textbf{C} for majority spin, (f) band \textbf{A} for minority spin, (g) band \textbf{B} for minority spin and (h) band  \textbf{C} for minority spin. Superlattice SrCeF$_2$ Bi$_2$ S$_4$ Pd$_4$: (j) crystal structure and (k) spin resolved electronic density of states.}
\label{Figure 1}
\end{figure}

\textbf{3. Theoretical Calculations: }\\
Within an LDA+$U$ treatment of electronic structure, Sr$_{0.5}$Ce$_{0.5}$FBiS$_2$ (crystal structure is shown in Figure 3(a)) is metallic in nature as seen in a nonzero electronic density of states at the Fermi level (see Figure 3(b)). The electronic states in the energy range between -7.8 eV and -6 eV are composed of F 2$p$ orbitals. S 3$p$ and Ce 4$f$ orbitals contribute to the density of states in the energy range between -6 eV and -1.5 eV constituting valence band similar to that in oxides. The states at the Fermi level in the conduction band are constituted of Bi 6$p$, Ce 4$f$, 5$d$ and Sr 5$s$ orbitals contribute to the conduction band. Due to magnetic exchange coupling, occupied up spin and unoccupied down spin states with Ce 4f orbital character are separated by 4 eV. One valence electron of Ce$^{3+}$ ion occupies the lowest energy $f$ orbital with up spin and contributes a local magnetic moment of 1 $\mu_B$ (per Ce$^{3+}$), which is more than its experimental value of $0.88$ $\mu_B$ at 2.5 K. We find three bands crossing the Fermi level (See spin resolved band structure in supplementary Figure S4(a) and  Figure S4(b)), and our estimation of the corresponding  Fermi velocities (see TABLE I) are in the range  4-6 $\times$ 10$^5$ m/s.

The Fermi surfaces associated with the three bands are all electron-like, and involve Bi 6$p$ orbitals (see Figure 3). The elliptical part of Fermi surface (centered at $\Gamma$ point of the Brillouin zone) associated with band A (see Figure 3(f)) of the minority spin distinguishes itself from that of the majority spin. Fermi surfaces associated with bands B and C with spin up and down are similar. The nonzero nesting wave vectors connecting the parallel flat parts of the spin resolved Fermi surfaces are q$_A$=0.392 $\mbox{\AA{}}^{-1}$ for band A and q$_B$=0.506 $\mbox{\AA{}}^{-1}$ for band B. Through a strong electron-phonon coupling, they are responsible for charge density waves as competing instabilities in the system. This is consistent with the observed pseudogap-like feature in the experimental data at higher temperatures. In addition, a soft Raman active mode of A$_{1g}$ symmetry with frequency 46 cm$^{-1}$ (See Figure S5(a)) involving the nonzero eigen-displacement of Bi and S atoms exhibits a strong coupling with electrons ($\lambda$=0.34), which can be understood from the fact that electronic density of states at the Fermi level arises from Bi 6$p$ orbitals.

\setlength{\tabcolsep}{2pt}
\begin{table}[ht] \caption{Fermi velocities in the bands crossing Fermi level of bulk Sr$_{0.5}$Ce$_{0.5}$FBiS$_2$.} \centering
\begin{tabular}{c c c c}
\hline \hline
\rule{0pt}{5ex}\parbox[b]{2cm}{Up spin\\band} & $v_F$\,(10$^5$ m/s) & \rule{0pt}{5ex}\parbox[b]{2cm}{Down spin\\band} & $v_F$\,(10$^5$ m/s) \\

\hline
A & 5.96 & A & 5.63\\
B & 5.75 & B & 5.41\\
C & 4.31 & C & 4.19\\

\hline \hline
\end{tabular} \label{table1}
\end{table}

The idea that the superconducting critical temperature here is suppressed by the fluctuation of the phase of the complex superconduting order parameter is also supported by the observation of a two-fold enhancement of the critical temperature formed under the point-contacts (5 K) than the bulk critical temperature (2.5 K) of SrCeF$_2$Bi$_2$S$_4$Pd$_4$. In order to understand this, we examine its interaction with Pd through determination of electronic structure of SrCeF$_2$Bi$_2$S$_4$Pd$_4$ superlattice, constructed by including two layers consisting of four Pd atoms on top of Bi-S layer along $c$ direction (See Figure 3(j)). 
 Metallic nature of Sr$_{0.5}$Ce$_{0.5}$FBiS$_2$ is retained even after its interaction with two atomic layers of Pd. However, this interaction perturbs the energy levels of all the atoms. The states in the energy range between -7.9 eV and -5 eV associated with F 2$p$ orbitals get shifted down in energy with respect to E$_F$. Nonzero density of states at the Fermi level arises from Pd 4$d$, S 3$p$ and Bi 6$p$ (See Figure 3(k)) orbitals, reflecting a strong hybridization between Pd 4$d$ and S 3$p$ orbitals which is not present in the electronic structure of the bulk. Local magnetic moment per cell (1 $\mu_B$) remains unchanged as there is no contribution coming from Pd 4$d$ orbitals to magnetic moment. We notice interaction with Pd leads to an increase in the density of states at the Fermi level (N(0)) of the superlattice by 78\% compared to bulk. 
More importantly, weaker nesting of multi-band Fermi surface of the superlattice (see Figure S6) compared to the bulk suggests suppression of the charge density wave instability resulting in  
enhancement of the superconducting transition temperature under the point-contacts. We note that the increase in density of states at the Fermi level is expected to make the phase of the order parameter stiffer against fluctuations. Depending on the order of shift in stiffness the critical temperature may increase up to a maximum of the mean field critical temperature. This idea was discussed in detail in the past in the context of high T$_c$ cuprates \cite{Phase}. This further supports the claim that quantum fluctuations of the phase of the complex superconducting order parameter caused by competing orders lead to the formation of a pseudogap in Sr$_{0.5}$Ce$_{0.5}$FBiS$_2$.

\textbf{4. conclusions:}\\
In conclusion, we have obtained a signature of a normal state pseudogap-like feature in the BiS$_2$ based ferromagnetic superconductor Sr$_{0.5}$Ce$_{0.5}$FBiS$_2$. From first-principles calculations, we have shown that multiple competing orders are supported by the electronic structure of Sr$_{0.5}$Ce$_{0.5}$FBiS$_2$ that might cause the suppression of superconducting $T_c$ through phase fluctuations leading to a pseudo-gap like feature. We have also observed a two fold enhancement of T$_c$ when a Pd-point-contact is formed on the surface of Sr$_{0.5}$Ce$_{0.5}$FBiS$_2$. We attribute this enhancement to the relative increase in the phase stiffness resulting from suppression of competing charge density wave instabilities due to weaker nesting of the Fermi surface.
\textbf{Acknowledgments:}\\
The authors at IISER Mohali acknowledge financial support from  DST Nanomission through grant number SR/NM/NS-1249/2013(G) and DST Ramanujan fellowship. The authors at IIT Delhi thank Prof. V. P. S. Awana for bulk resistivity measurements. AKG and UVW thank DST for financial support. MS and GST thank CSIR for a research fellowship. AP is thankful for the research fellowship from the Department of Science and Technology and TUE - CMS, JNCASR for computational resources.\\


\textbf{References:}


\begin{thebibliography}{100}


\bibitem {Bardeen} J. Bardeen, L. N. Cooper, J. R. Schrieffer, Phys. Rev. \textbf{106}, 162 (1957).
\bibitem{Berk} N. F. Berk and J. R. Schrieffer, Phys. Rev. Lett. \textbf{17}, 433 (1966).
\bibitem{Al} Alexandre Buzdin, Nature Materials \textbf{3}, 751 - 752 (2004).
\bibitem{Ber} Bernd Lorenz, Ching-Wu Chu,  Nature
Materials. \textbf{4} ,  516 - 517 (2005).
\bibitem{n.d} N. D. Mathur \emph{et al.}, Nature (London). \textbf{394}, 39-43 (1998).
\bibitem{r.e} R. E. Baumbach, V. A. Sidorov, Xin Lu, Nirmal J Ghimire, F. Ronning, Brian L Scott, Phys. Rev. B \textbf{89},   094408  (2014).
\bibitem{m.s} M. S. Torikachvili, S. L. Bud'ko, Ni Ni, P. C. Canfield, Phys. Rev. Lett. \textbf{101}, 057006   (2008).

\bibitem{yong} Yongkang Luo, Han Han, Shuai Jiang, Xiao Lin, Yuke Li, Jianhui Dai, Guanghan Cao, Zhu-an Xu,
Phys. Rev. B \textbf{83}, 054501 (2011).
\bibitem{Li} Lin Li \emph{et al.}, Phys. Rev. B \textbf{91}, 014508 (2015).

\bibitem{Ren} Zhi Ren, Qian Tao, Shuai Jiang, Chunmu Feng, Cao Wang, Jianhui Dai, Guanghan Cao, Zhu’an Xu, Phys. Rev. Lett. \textbf{102}, 137002  (2009).

\bibitem{J.Lee} J. Lee \emph{et al.}, Phys. Rev. B \textbf{90}, 224410 (2014).

\bibitem{s.s} S. S. Saxena \emph{et al.}, Nature (London) \textbf{406}, 587-592 (2000).
\bibitem{d} D. Aoki, A. Huxley, E. Ressouche, D. Braithwaite, J. Flouquet, J. Brison, E. Lhotel, and C. Paulsen, Nature (London) \textbf{413}, 613-616 (2001).

\bibitem{Jiangping} Jiangping Hu, Hong Ding, Scientific Reports \textbf{2}, 381 ( 2012).

\bibitem{Xing} Jie Xing, Sheng Li, Xiaxin Ding, Huan Yang, Hai-Hu Wen, Phys. Rev. B. \textbf{86}, 214518 (2012).

\bibitem{Usui} Hidetomo Usui, Katsuhiro Suzuki, Kazuhiko Kuroki, Phys. Rev. B. \textbf{86}, 220501(R) (2012).

\bibitem{Yild} T. Yildirim, Phys. Rev. B \textbf{87}, 020506 (2013).

\bibitem{Yoshkazu} Yoshikazu Mizuguchi \emph{et al.}, Phys. Rev. B \textbf{86}, 220510 (R) (2012).
\bibitem{Xi} Xi Lin \emph{et al.}, Phys. Rev. B. \textbf{87}, 020504(R) (2013).

\bibitem {Sugi}  T. Sugimoto, B. Joseph, E. Paris, A. Iadecola, T. Mizokawa, S. Demura, Y. Mizuguchi, Y. Takano, N. L. Saini, Phys. Rev. B. \textbf{89}, 201117(R) (2014).

\bibitem{Zeng} L. K. Zeng \emph{et al.}, Phys. Rev. B. \textbf{90}, 054512 (2014).
\bibitem{H}  H. C. Lei, K. F. Wang, M. Abeykoon, E. S. Bozin and C. Petrovic, Inorg. Chem. (Washington, DC, US) \textbf{52}(18), 10685 (2013).
\bibitem{Y}  Yoshikazu Mizuguchi, Satoshi Demura, Keita Deguchi, Yoshihiko Takano, Hiroshi Fujihisa, Yoshito Gotoh, Hiroki Izawa,  Osuke Miura-, Journal of the Physical Society of Japan \textbf{81},  114725 (2012).

\bibitem {S}  Satoshi Demura \emph{et al.}, Journal of the Physical Society of Japan \textbf{82}, 033708 (2013).

\bibitem{M.d} M. D. Johannes, I. I. Mazin, Phys. Rev. B \textbf{77}, 165135 (2008).
\bibitem{wan} Xiangang Wan, Hang-Chen Ding, Sergey Y. Savrasov, Chun-Gang Duan, Phys. Rev. B. \textbf{87}, 115124 (2013).
\bibitem{Lamura}  G. Lamura \emph{et al.}, Phys. Rev. B. \textbf{88}, 180509(R) (2013).

\bibitem {Ye} Z. R. Ye \emph{et al.}, Phys. Rev. B. \textbf{90}, 045116 (2014).
\bibitem {Baumb} F. Baumberger \emph{et al.}, Phys. Rev. Lett. \textbf{96}, 107601 (2006).
\bibitem {G.B} G. B. Martins,  A. Moreo, and E. Dagotto, Phys. Rev. B \textbf{87},  081102(R)  (2013).
\bibitem {Lee} J. Lee et at., Phys. Rev. B \textbf{87}, 205134 (2013).

\bibitem {Goutam} Goutam Sheet, Manan Mehta,D. A. Dikin, S. Lee, C. W. Bark, J. Jiang, J. D. Weiss, E. E. Hellstrom, M. S. Rzchowski, C. B. Eom, and V. Chandrasekhar,  Phys. Rev. Lett. \textbf{105}, 167003 (2010).

\bibitem{Davis}Jhinhwan Lee, K. Fujita, A. R. Schmidt, Chung Koo Kim, H. Eisaki, S. Uchida, J. C. Davis, Science \textbf{325}, 1099 (2009).
 
\bibitem{Pratap} Mintu Mondal, Anand Kamlapure, Madhavi Chand, Garima Saraswat, Sanjeev Kumar, John Jesudasan, L. Benfatto, Vikram Tripathi, and Pratap Raychaudhuri, Phys. Rev. Lett. \textbf{106}, 047001 (2011).

\bibitem{Emery} V. J. Emery, S. A. Kivelson, Nature \textbf{374}, 434 (1994).
\bibitem {sinchenko} A. A. Sinchenko, P. Monceau, Phys. Rev. B \textbf{67}, 125117 (2003).

\bibitem{Phase} Erez Berg, Dror Orgad, and Steven A. Kivelson
Phys. Rev. B \textbf{78}, 094509 (2008).

\end{thebibliography}
\end{document}